\documentclass[a4paper,fleqn,usenatbib]{mnras}
\usepackage[T1]{fontenc}
\usepackage{ae,aecompl}
\usepackage{graphicx}	
\usepackage{amsmath}	
\usepackage{amssymb}	

\def\eq#1{\begin{equation} #1 \end{equation}}
\def\E#1{\hbox{$10^{#1}$}}
\def\about  {\hbox{$\sim$}}
\def\x      {\hbox{$\times$}}
\def\A      {\AA}
\def\Mo     {\hbox{$M_{\sun}$}}
\def\erg    {\hbox{erg\,s$^{-1}$}}
\def\cs     {\hbox{cm$^{-2}$}}
\def\kms    {\hbox{km\,s$^{-1}$}}

\def\v{\m{166}}

\def\vK     {\hbox{$\v_{\rm K}$}}
\def\vd     {\hbox{$\v_{\rm Kd}$}}
\def\Rd     {\hbox{$R_{\rm d}$}}
\def\Mdot   {\hbox{$\dot M$}}
\def\Mw     {\hbox{$\dot M_{\rm w}$}}
\def\Ncrit  {\hbox{$N_{\rm crit}$}}
\def\Nw     {\hbox{$N_{\rm w}$}}
\def\Nmin   {\hbox{$N_{\rm H,min}$}}
\def\NH     {\hbox{$N_{\rm H}^c$}}
\def\NR     {\hbox{$N_R$}}
\def\No     {\hbox{${\cal N}_0$}}
\def\mH     {\hbox{$m_{\rm p}$}}
\def\Lmin   {\hbox{$L_{\rm min}$}}
\def\Ledd   {\hbox{$L_{\rm Edd}$}}
\def\LEH    {\hbox{$L_{\rm EH}$}}
\def\Lmm    {\hbox{$L_{\rm min>}$}}
\def\emm    {\hbox{$\epsilon_{\rm m}$}}
\def\Ha     {\hbox{${\rm H\alpha}$}}
\def\LbHa   {\hbox{$L_{\rm bH\alpha}$}}
\def\Hb     {\hbox{${\rm H\beta}$}}

\def\NRB    {\hbox{$N_R^{\rm BLR}$}}
\def\NRT    {\hbox{$N_R^{\rm TOR}$}}
\def\MwB    {\hbox{$\dot M_{\rm w}^{\rm BLR}$}}
\def\MwT    {\hbox{$\dot M_{\rm w}^{\rm TOR}$}}
\def\IB     {\hbox{$I^{\rm BLR}$}}
\def\IT     {\hbox{$I^{\rm TOR}$}}
\def\ybar   {\hbox{${\bar y}$}}

\title[High-Luminosity True Type 2 AGN]
            {Disk Outflows and High-Luminosity True Type 2 AGN}

\author[Elitzur, \& Netzer]
       {Moshe Elitzur$^{1,2}$ and Hagai Netzer$^{3}$\\
   $^1$Astronomy Department,
       University of California,
       Berkeley, CA 94720-3411,
       USA\\
   $^2$Department of Physics and Astronomy,
       University of Kentucky,
       Lexington, KY 40506-0055,
       USA\\
   $^3$School of Physics and Astronomy,
       Tel Aviv University,
       Tel Aviv 69978,
       Israel
       }

\date{Submitted August 13, 2015; revised January 20, 2016; accepted March 15, 2016}
\pubyear{2016}

\begin{document}
\label{firstpage} \pagerange{\pageref{firstpage}--\pageref{lastpage}}
\maketitle

\begin{abstract}
The absence of intrinsic broad line emission has been {reported in a number
of active galactic nuclei (AGN), including some with high Eddington ratios}.
Such ``true type 2 AGN" are inherent to the disk-wind scenario for the broad
line region: Broad line emission requires a minimal column density, implying
a minimal outflow rate and thus a minimal accretion rate. Here we perform a
detailed analysis of the consequences of mass conservation in the process of
accretion through a central disk. {The resulting constraints on luminosity
are consistent with all the cases where claimed detections of true type 2 AGN
pass stringent criteria, and predict that intrinsic broad line emission can
disappear at luminosities as high as \about 4\x\E{46}\,\erg\ and any
Eddington ratio, though more detections can be expected at Eddington ratios
below \about 1\%.} Our results are applicable to every disk outflow model,
whatever its details and whether clumpy or smooth, irrespective of the wind
structure and its underlying dynamics. { While other factors, such as changes
in spectral energy distribution or covering factor, can affect the
intensities of broad emission lines, within this scenario they can only
produce true type 2 AGN of higher luminosity then those prescribed by mass
conservation.}

\end{abstract}


\section{Introduction}

Although AGN unification is supported by a large body of evidence, it is still
incomplete in several ways. In the standard unification scenario, all AGN are
intrinsically the same, except for their luminosity, and produce the same broad
emission line spectrum. The only reason we do not observe the broad emission
lines in type 2 AGN is our viewing angle with respect to the toroidal
obscuration that surrounds the central engine \citep[e.g.][]{Ski93}. However,
the structure of the AGN environment must evolve with accretion rate and
luminosity---obviously the AGN disappears altogether once the accretion rate
drops to zero. The question is whether the evolution induced by the decline of
accretion rate produces discernible effects before that final point is reached;
that would imply that all AGN are in fact not intrinsically the same.

{An increasing number of AGN are reportedly missing} broad emission lines, or
the lines are extremely weak, even though there is little or no obscuration to
their central source. First noticed in low luminosity AGN with low Eddington
ratios \citep[see, among others,][]{Tran01, Tran03, Panessa02, Laor03}, this
phenomenon has now been {reported} also in high Eddington-ratio AGN
\citep{Ho12, Miniutti13}. Since the first {reported} objects of this type show
prominent narrow emission lines, they were referred to as ``true type 2 AGN''.
However, the unusual appearance is related more to the strength of the broad
emission lines relative to the non-stellar continuum, and the above name (which
will be retained for the rest of the paper) will be used to describe AGN with
weak broad emission lines, whatever the narrow line strength.

{Conclusively establishing the absence of emission features is always
difficult. Indeed, \cite{Stern12a, Stern12b} have noted that the observed upper
limits on broad \Ha\ emission in some previously claimed true type 2 AGN are
insufficient to establish decisively the true nature of these sources. This
ambiguity does not mean that broad \Ha\ must be there, only that the data are
not yet conclusive in demonstrating its complete absence. Stern \& Laor also
noted that there are sources where the absence of broad \Ha\ emission is
established with such stringent upper limits, well below the level expected
from their X-ray luminosities, that they indeed appear to be true type 2 AGN.
Another serious difficulty is time variability. Some AGN, dubbed ``changing
look quasars", display temporal variations in broad line strengths that cannot
be attributed to transient obscuration and must be related to a large sudden
drop in accretion rate (see \citealt{LaMassa15, MacLeod15}, and references
therein). While some of the suggested cases of true type 2 AGN undoubtedly
reflect such temporary reductions in broad \Ha\ strength, multi-epoch
observations preclude this possibility in a number of others. On the basis of
current observational evidence, here we take the view that the true type 2
phenomenon merits additional serious considerations and seek a general
theoretical framework to explain it.}

Observations suggest that the broad-line region (BLR) and the toroidal
obscuration region (TOR, a.k.a. ``the torus") may be the inner and outer zones,
respectively, in a single, continuous distribution of material whose
composition undergoes a change at the dust sublimation radius (see
\citealt{Elitzur08}, and references therein). Such a structure arises naturally
in the disk-wind scenario, where the source of material to the BLR and the TOR
is mass outflow from the surface of a disk that extends to distances that are
an order of magnitude larger than a ``typical" BLR size (measured by
reverberation mapping to be \about \E3--\E4\ gravitational radii). The
immediate prediction of this scenario \citep{Elitzur_Shlosman} is that  at low
accretion rates, the mass outflow rate decreases too and the wind radial column
density eventually drops below the minimum required to produce detectable broad
line emission in the dust-free zone and continuum obscuration in the dusty
zone. In this scenario, true type 2 AGN inevitably emerge at low accretion
rates.

\citet{EH09} presented a preliminary analysis of mass conservation in the
disk-wind scenario at low accretion rates. They argued that the disappearance
of the BLR and TOR should occur at bolometric luminosities that obey $L \propto
M^{2/3}$, where $M$ is the black-hole mass, with a proportionality coefficient
that varies among sources. While that study focused on the low-luminosity end,
here we present a more complete analysis and examine in detail whether mass
conservation and the constraints it implies can also explain the {reported}
high-luminosity true type 2 AGN.

\section{Constraints on Broad Line Emission}
\label{sec:basics}

Detectable broad line emission requires a minimal column density, \Nmin, to
produce radial stratification with photoionization and recombination events
occurring at a sufficient rate. The fundamental quantity \Nmin\ that sets the
threshold for detectable emission lines is determined by physical processes and
atomic constants, and is largely independent of the AGN detailed structure.
Following the results of detailed photoionization calculations
\citep[e.g.,][]{Netzer13}, we adopt \Nmin\ = 5\x\E{21} \cs. Denote
\eq{\label{eq:NR}
   \NR = \int n(R) dR
}
the column density along a radial ray close to the AGN equatorial plane, where
$R$ is axial radius and $n$ is the gas density. Since the column is expected to
decrease, or at least not to increase, away from the equator toward the poles,
no other radial column is larger than \NR. Therefore, to produce a BLR the
column \NR\ must exceed \Nmin, namely,
\eq{\label{eq:constraint1}
   \NR > \Nmin
}
This is a fundamental constraint that must be obeyed by all broad-line emitting
AGN. It involves no assumptions about the BLR detailed structure or dynamics.

\subsection{Mass Conservation Constraint}

While the column \Nmin\ is controlled by basic physical processes, \NR\ is a
specific property of each AGN and cannot be calculated without some assumptions
about the structure of its BLR. Here we invoke the disk-wind scenario for the
BLR and derive \NR\ from mass continuity for the outflow.

In the disk-wind scenario, the BLR corresponds to a portion of the wind ejected
from a finite annular segment of the disk whose outer radius is set by dust
sublimation. Denote by \Mw\ the overall mass outflow rate from that segment of
the disk. Accounting for emission from both faces of a
disk\footnote{\cite{EH09} is missing a factor of 2 for emission from both
sides.}, the mass conservation relation for a smooth-density disk outflow is
\eq{\label{eq:Mw0}
    \Mw = 4\pi \mH \int\!n\v_z RdR,
}
where \mH\ is the proton mass, $\v_z$ is the outflow velocity vertical
component at the disk surface and $R$ and $n$ as in eq.\ \ref{eq:NR}. We now
introduce characteristic scales for the variables. For $R$ we use the dust
sublimation radius \Rd\ so that dimensionless distance from the axis is
\eq{
   y = \frac{R}{\Rd}
}
and the origin of the BLR outflow corresponds to disk radii $y \le 1$ (the TOR
has $y \ge 1$). We write the launch velocity as $\v_z(R) = \vK(R)f(R)$, where
$\vK(R)$ is the local Keplerian velocity and $f\ (< 1)$ a dimensionless
profile, and introduce
\eq{
   \vd = \vK(\Rd) = \left(\frac{GM}{\Rd}\right)^{1/2}.
}
Then the outflow launch velocity becomes
\eq{
    \v_z(y) = \vd\frac{f(y)}{y^{1/2}}.
}
With \NR\ from eq.\ \ref{eq:NR}, we extract from the density $n$ its
dimensionless radial profile $\eta$ through
\eq{\label{eq:eta}
    n(y) = \frac{\NR}{\Rd}\eta(y),
    \qquad \hbox{so that} \quad
    \int\eta(y)dy = 1.
}
The profile $\eta$, normalized to unit integral, describes the functional form
of  the  density radial variation, e.g., a power law, etc. (see \S\ref{sec:I}
below). With these definitions, eq.\ \ref{eq:Mw0} becomes
\eq{\label{eq:Mw}
  \Mw = 4\pi\mH\,\Rd \vd \NR\, I,
    \qquad \hbox{where} \quad
    I = \int f\eta y^{1/2}dy.
}
This is the mass conservation relation in terms of the radial column \NR\
through the disk outflow. Inserting \NR\ from this result into eq.\
\ref{eq:constraint1} yields a constraint on the mass outflow rate in broad-line
emitting AGN. This constraint can be cast in the convenient form
\eq{\label{eq:Nw}
   \Nw > I\Nmin,
   \quad \hbox{where} \ \
   \Nw = \frac{\Mw}{4\pi\mH \Rd \vd}
}
is a characteristic column density formed out of the outflow parameters (see
\citealt{EHT14}, which used the notation \Ncrit\ for this quantity). This is
the BLR constraint (eq.\ \ref{eq:constraint1}) in the disk-wind scenario,
expressed in terms of global properties of the system. All detailed information
about the specific structure of a given AGN, i.e., the functional forms of its
density and velocity radial distribution profiles, is contained in a single
factor, the dimensionless integral $I$.

A clumpy disk-outflow yields the exact same result. Denote by $M_c$ the mass of
a single cloud, $\v_c$ its vertical launch velocity and $n_c$ the number
density of clouds (number of clouds per unit volume).  Integrating over the
disk area then gives $\Mw = 4\pi\int M_c n_c\v_c RdR$. If \NH\ is the column
density of a single cloud and $A_c$ its cross-sectional area then $M_c = \mH
\NH A_c$. Since $N_c \equiv n_cA_c$ is the number of clouds per unit length,
$\Mw = 4\pi \mH\int \NH N_c \v_c\, RdR$. Denote by $\No
\equiv \int N_c\,dR$ the (mean of the) total number of clouds along a radial
equatorial ray, then the total radial column at the base of the wind (just
above the disk surface) is \NR = \No\NH. Introducing $\eta = N_c(R)\Rd/\No$
yields back eq.\ \ref{eq:Mw} with the same definition of $f$; the only
difference is that now $\eta$ is the normalized radial distribution of the
number of clouds per unit length instead of gas density. Given this, there is
no need to distinguish between the smooth and clumpy components of the wind;
\NR\ is the total radial column density of the outflow at its origin above the
disk surface irrespective of clumpiness.

The constraint in eq.\ \ref{eq:Nw} on broad-line emitting AGN is a fundamental
property of the disk-wind scenario. As a direct consequence of mass
conservation (eq.\ \ref{eq:Mw0}), it is in essence a kinematic outcome of this
scenario that does not involve any assumptions about the wind dynamics, and is
applicable for both smooth and clumpy outflows.

\subsection{Luminosity Constraint}
\label{sec:L}

Equation \ref{eq:Nw} provides a general constraint on disk outflows in
broad-line emitting AGN. Unfortunately, this relationship has limited direct
utility since it involves the BLR mass outflow rate \Mw, which is not readily
measurable. For comparison with observations we wish to relate \Mw\ to the
bolometric luminosity $L$. Black-hole mass accretion at the rate \Mdot\
generates the luminosity $L = \epsilon\Mdot c^2$, where $\epsilon$ is the
radiative conversion efficiency. Since we are interested in the mass outflow
rate \Mw, we introduce the ratio $r = \Mdot/\Mw$ so that
\eq{\label{eq:L}
   L = \epsilon r\Mw c^2.
}
In this variation on the standard luminosity relation, the black-hole mass
accretion rate \Mdot\ is replaced by the BLR outflow rate \Mw, hence the
radiative efficiency is replaced by the product $\epsilon r$. With this, eq.\
\ref{eq:Mw} for the relation between \Mw\ and \NR\ becomes
\eq{\label{eq:L-NR}
      L = 4\pi\mH c^2\, \epsilon r I\, (GM\Rd)^{1/2} \NR.
}
This is the relation between the AGN luminosity and the radial column density
of the outflow at its origin above the disk. It is not yet in suitable form
since the luminosity enters also indirectly on the right hand side because of
the $L$-dependence of dust sublimation. The dust sublimation radius obeys the
relation $\Rd = b\,L_{45}^{1/2}$ pc, where $L_{45} = L/\E{45}\,\erg$ and $b$ is
a normalization factor that depends on the dust composition. \cite{AGN2}
considered a standard graphite-silicate dust mixture that yields $b$ = 0.43.
\cite{Mor12} suggest that, since only pure graphite grains survive in the torus
innermost part, $b \simeq 0.16$ should be used instead. This is in good
agreement with the most recent dust reverberation results of \cite{Koshida14},
and thus is the value we adopt here. Then eq.\ \ref{eq:L-NR} becomes
\eq{
   \NR = \frac{2.07\x\E{21}}{\epsilon r I}
         \left(\frac{L_{45}}{M_7^{2/3}}\right)^{3/4} \cs,
}
where $M_7 = M/10^7\Mo$. This is the final relation for radial column density
in terms of mass and luminosity in all disk outflows. With it, the broad-line
emission constraint $\NR
> \Nmin$ (eq.\ \ref{eq:constraint1})  becomes
\eq{\label{eq:Lmin0}
 L > \Lmin,
}
where
\eq{\label{eq:Lmin}
   \Lmin = \Lambda\,M_7^{2/3},
   \hskip0.12in \hbox{and} \quad
   \Lambda = 3.25\x\E{45}\,(\epsilon\,r I)^{4/3}\ \erg.
}
This is the minimal luminosity for the existence of an observable BLR in the
disk-wind scenario \citep[cf][]{EH09}. Objects whose luminosity falls below
this limit do not have a visible BLR and are classified as true type 2 AGN. The
relation between \Lmin\ and the Eddington luminosity, $\Ledd = 1.26\x\E{45}M_7\
\erg$, is
\eq{\label{eq:Lmin-Edd}
   \frac{\Lmin}{\Ledd} = 2.58\left[\frac{(\epsilon\,r I)^4}{M_7}\right]^{1/3}
}
Systems with masses $M_7 < 17.2\,(\epsilon r I)^4$ will have $\Ledd < \Lmin$,
thus they will necessarily be true type 2 AGN if they have sub-Eddington
luminosities.

\section{Numerical Estimates}

Equation \ref{eq:Lmin0} provides the fundamental constraint on broad line
emission in the disk-wind scenario, imposing a lower limit on AGN luminosity
that varies with the black-hole mass. Because each of the three parameters $I$,
$r$ and $\epsilon$ can vary from source to source, $\Lambda$ is expected to
vary among AGN (eq.\ \ref{eq:Lmin}) and systems with the same $M$ could still
be subjected to different lower limits on broad line emission. Thus we need to
find the variation range of the luminosity scale $\Lambda$.

The lower end of this range can be estimated directly from observations. From
the Palomar AGN sample, \cite{EH09} find that all broad line emission
disappears at luminosities lower than
\eq{\label{eq:LEH}
   \LEH = 4.7\x\E{39}\,M_7^{2/3}   \quad \erg.
}
Unification then implies that all AGN below this boundary are true type~2
(otherwise there would have been some type 1 counterparts). Since it is based
on observations of low-luminosity AGN, this limit is somewhat uncertain, owing
to the large uncertainties in determining bolometric luminosities in this
regime. The \LEH\ limit conforms to the luminosity constraint of the disk-wind
scenario (eq.\ \ref{eq:Lmin}), implying a minimum for $\Lambda$ of
\eq{\label{eq:Lam_min}
   \Lambda_{\min} \simeq 4.7\x\E{39}\,\erg.
}
This minimum is reached when all three parameters $I$, $r$ and $\epsilon$ have
their smallest values. When each has its highest value, the maximum of
$\Lambda$ is obtained. We now estimate one by one the maxima of these three
parameters.

\subsection{The $I$-factor}
\label{sec:I}

To estimate the integral $I$ (eq.\  \ref{eq:Mw}) we need the ratio of outflow
launch velocity to local Keplerian velocity, $f = \v_z(R)/\vK(R)$. The launch
velocity is expected to be comparable to the local turbulent velocity (the
critical sonic-point condition), thus the quantity needed is the ratio of
turbulent velocity to local Keplerian velocity in the disk. The most direct
estimate for this ratio  comes from high-resolution H$_2$O maser observations.
Such observations of the nuclear disk in NGC~3079 show that the velocity
dispersion is \about\ 14 \kms\ over a small region of strong emission where the
Keplerian velocity is 110 \kms\ \citep{Kondratko05}, suggesting that in the
maser region $f$ is probably of order \about\,0.1. Given the lack of
information about the turbulent velocity close to the inner disk producing the
BLR wind, we assume that the value of f in this region is similar to that
derived from the observations further out in the maser region; that is, we make
the assumption that at every point in the disk, the turbulent velocity is a
fixed fraction, $f$ \about\ 10\%, of the local Keplerian velocity. Then
\eq{\label{eq:J}
   I \simeq 0.1J,
   \qquad \hbox{where} \quad
   J = \int \eta y^{1/2}dy
}
and $\eta$ is the profile of the radial variation of density, properly
normalized (eq.\ \ref{eq:eta}).

\begin{figure}
  \centering
\includegraphics[width=\hsize]{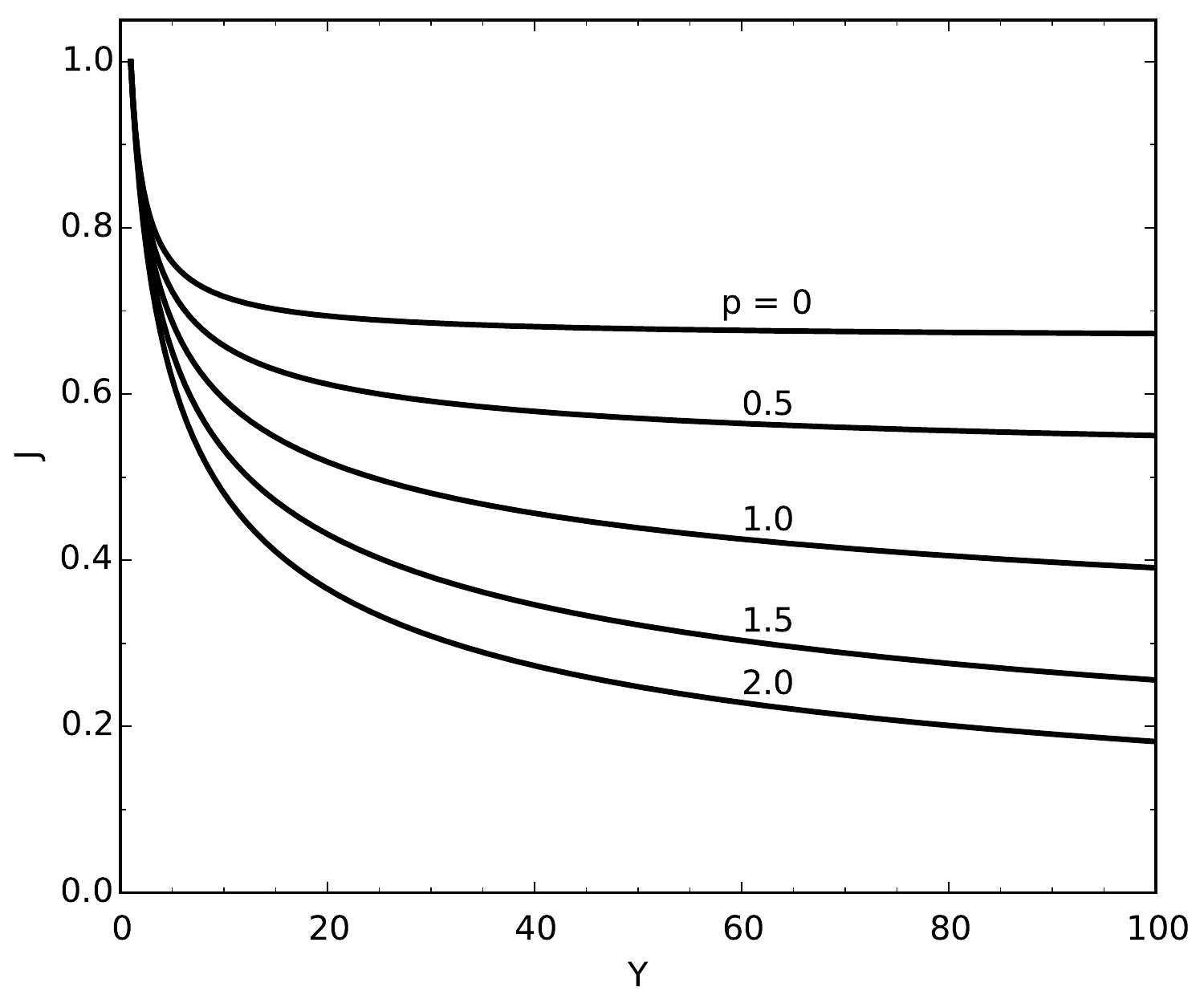}

\caption{The integral $J$ (eq.\ \ref{eq:J}) for power-law radial profiles $\eta
\propto y^{-p}$ for various values of $p$ as a function of $Y = R_{\rm
out}/R_{\rm in}$, the relative radial thickness; the BLR radial integration
range is $1/Y \le y \le 1$. Note that $J \to 1$ when $Y \to 1$ because of the
normalization of $\eta$ (eq.\ \ref{eq:eta}).
}\label{fig:Ip}
\end{figure}

Without a detailed model for the BLR dynamics, the functional form of $\eta$
remains unknown. However, we can still place reasonable limits on the integral
$J$ from various power-laws $\eta \propto y^{-p}$, for which the integration is
immediate. Figure \ref{fig:Ip} shows the variation of $J$ with relative radial
thickness $Y$ (ratio of outer-to-inner radius) for representative values of
$p$. The full radial extent of the BLR is expected to be in the range $R_{\rm
out}/R_{\rm in} \sim 50-100$, and figure \ref{fig:Ip} shows that $J$ hardly
varies with $Y$ in this range. The bulk of the variation comes from the
dependence on $p$; $J$ varies from \about~0.2 for $p = 2$ to \about~0.7 for $p
= 0$. These values bracket the likely range of $J$ for actual $\eta$ profiles.
Therefore
\eq{\label{eq:I}
   I_{\max} \simeq 0.07,
}
a maximum reached for flatter radial density distributions. The figure also
shows that the assumption of constant $f$ should not have a marked impact on
the outcome---a small radial variation of $f(R)$ could simply be absorbed into
the power law.

\subsection{The radiative efficiency $\epsilon$}
\label{sec:epsilon}

Beginning with the early 1970's, the theory of disk accretion has been
investigated in numerous studies. In the standard theory, the mass accretion
rate through the disk is constant. Outflows invalidate this assumption, and
recent studies by \cite{SloneNetzer12} and \cite{Laor14} were the first to
investigate their effect on the observed disk spectra. We assume that most of
the mass outflow takes place at distances that are typical of the BLR, shown by
reverberation mapping to be \about \E3--\E4\,$GM/c^2$, well outside the main
continuum producing region, which in standard thin accretion models is of order
50--100\,$GM/c^2$. This allows us to use the standard disk theory to estimate
the bolometric luminosity.

For geometrically-thin optically-thick disks the radiative efficiency ranges
from 0.038 for a retrograde disk, to 0.057 for stationary black hole (BH) to
0.32 for maximally rotating BH with spin parameter of 0.998 \citep{Shakura73}.
When the effects of radiation capture by the BH are accounted for, $\epsilon$
at maximal spinning is reduced to 0.3 \citep{Thorne74}. Slim accretion disks
were suggested to have smaller efficiency (\citealt{Wang14} and references
therein), but this is not yet confirmed by numerical simulations
\citep[e.g.][]{Sadowski15}. Recent observational studies provide support for
the radiatively efficient solutions for accretion through thin disks. From
detailed analysis of spectral properties, \cite{Davis11} estimated the values
of $\epsilon$ for all PG quasars. The majority of the sources are consistent
with the theoretical solutions for radiatively efficient accretion through thin
disks, with $\epsilon$ falling between \about \hbox{6--30\%}. A more detailed
analysis of 39 luminous AGN at $z \simeq 1.55$, by \cite{Capellupo15,
Capellupo16}, illustrated the very good agreement between the predicted thin
disk SED and the observations of all these objects. Given this, we can
reasonably take for $\epsilon$ the upper limit
\eq{\label{eq:epsilon}
   \epsilon_{\max} \simeq 0.3,
}
the radiative efficiency of a maximally spinning accreting black hole.

\subsection{The $r$-ratio}
\label{sec:r}

This is the ratio $r = \Mdot/\Mw$ of mass accretion into the black hole to mass
outflow from the BLR (\S\ref{sec:L}). Because of the disk outflow, not all the
mass that enters the BLR finds its way to the black hole. Denote by $\Mdot_{\rm
outer}$ the rate at which mass enters the BLR through its outer disk boundary
(the sublimation radius \Rd) and by $\gamma = \Mw/\Mdot_{\rm outer}$ the
fraction of that accreted material lost to the wind. Then the rate of radial
inflow from the BLR inner boundary toward the black hole is $\Mdot_{\rm
outer}(1 - \gamma)$. As just noted (\S\ref{sec:epsilon}), the radial mass
inflow rate through the disk can be taken as constant inside the BLR inner
radius, thus the black-hole mass accretion rate is $\Mdot = \Mdot_{\rm outer}(1
- \gamma)$ and
\eq{\label{eq:r-gamma}
    r = \frac{1 - \gamma}{\gamma}.
}
The largest $r$ corresponds to the smallest $\gamma$, obtained when the
fractional mass carried away by the wind is at minimum. However, the limit
\hbox{$\gamma \to 0$} implies shutting off the outflow altogether, as it
carries a smaller and smaller fraction of the accreted mass; that in itself
would turn off broad line emission in the context of the disk-wind scenario
assumed here. For any relevance in this context and to have an impact on
observable AGN properties, the disk wind must carry some minimal fraction of
the accreted mass. Here we assume, somewhat arbitrarily, that for the
self-consistency of this scenario this fraction should be at least \about 10\%.
This requirement implies that $\gamma$ must be at least \about 0.1 in
meaningful disk-wind models of the BLR, leading to
\eq{\label{eq:r}
    r_{\max} \simeq 9.
}

\subsection{The range of $\Lambda$}

The minimal value of $\Lambda$, the proportionality coefficient in the
luminosity constraint on broad line emission (eq.\ \ref{eq:Lmin}), was
determined directly from observations in eq.\ \ref{eq:Lam_min}. Determining the
maximum requires a theoretical estimate. Combining our estimates for maximal
$I$ (eq.\ \ref{eq:I}), $\epsilon$ (eq.\ \ref{eq:epsilon}) and $r$ (eq.\
\ref{eq:r}) yields
\eq{\label{eq:Lam_max}
    \Lambda_{\max} \simeq 3.5\x\E{44}\ \erg
}
A large $\Lambda$ means that observable broad line emission requires a large
bolometric luminosity, increasing the range of luminosities that can lead to
true type 2 operation. An AGN is more likely to be a true type 2 if it has $I$,
$r$ and $\epsilon$ at the upper end of their ranges. A larger $I$ implies a
flatter density profile (fig.\ \ref{fig:Ip}): a larger fraction of the outflow
mass resides at larger radii, therefore the same minimal column is accompanied
by a larger \Mw, i.e., $L$. Larger $r$ imply that the outflow carries away a
smaller fraction of the mass (eq.\ \ref{eq:r-gamma}), requiring a larger \Mw,
i.e., $L$, in order to reach the minimal BLR column density. And a larger
$\epsilon$ means that the mass accretion rate corresponding to the minimal
column for a BLR will produce a higher luminosity. At the upper end of
$\Lambda$, an AGN with a maximally rotating BH ($\epsilon$ = 0.3) whose mass is
\E{10}\,\Mo\ will be true type 2 as long as it has $L < 3.5\x\E{46}\,\erg$, a
rather high luminosity threshold.

\begin{figure}
  \centering
\includegraphics[width=\hsize]{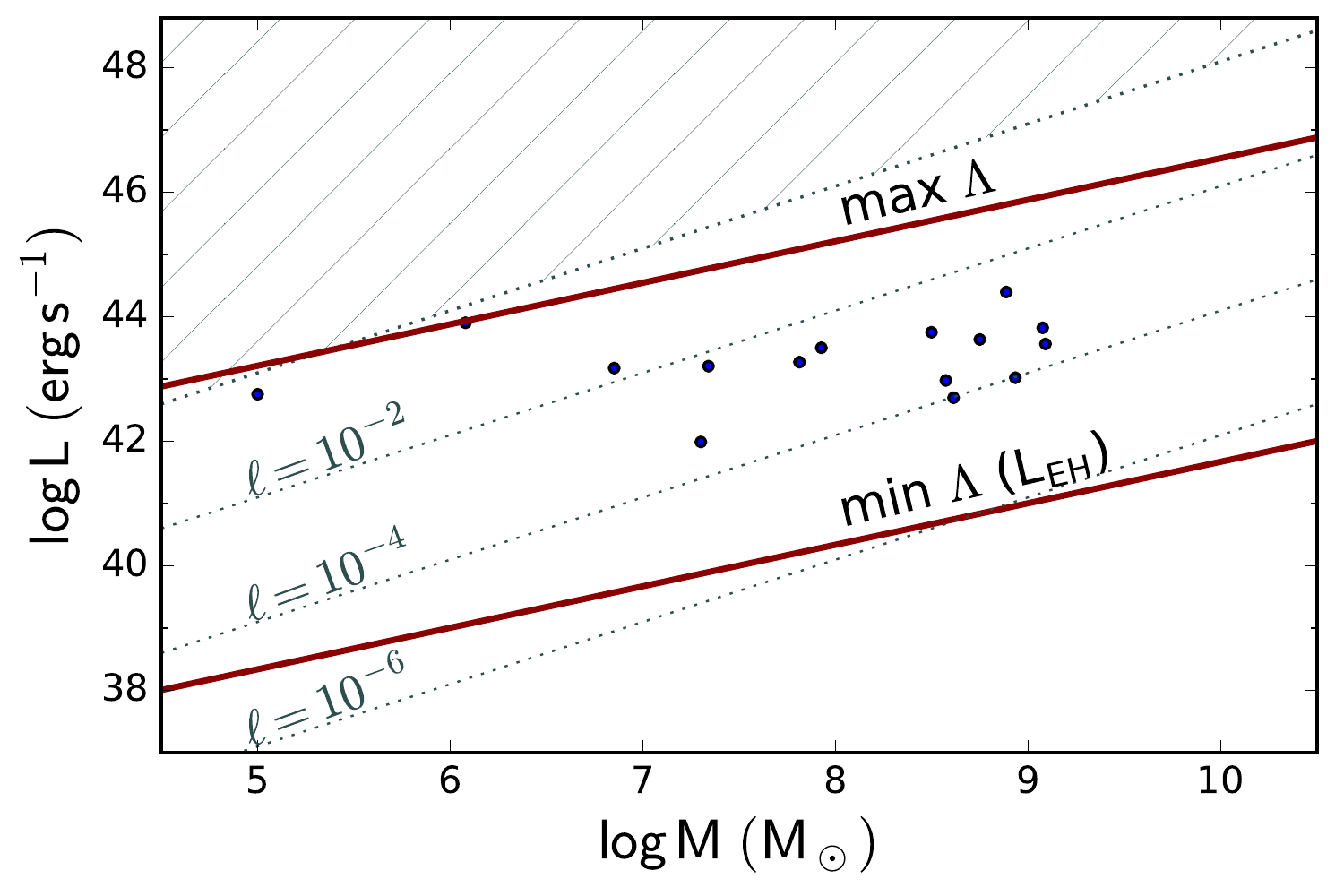}

\caption{Domains in the $L$--$M$ plane. The hatched region marks
super-Eddington luminosities; also shown are contours (gray dotted lines) for
several Eddington ratios $\ell = L/\Ledd$, as marked. The two red curves show
the low-luminosity boundary for broad line emission (eq. \ref{eq:Lmin}) for the
minimal and maximal values of $\Lambda$ (eqs.\ \ref{eq:Lam_min},
\ref{eq:Lam_max}), as marked. According to the disk-wind scenario all sources
below the low red boundary are true type 2 AGN, all sources above the upper one
are broad line emitters and in between they can be either. Dots are the data
from Table 1.
}\label{fig:L-M}
\end{figure}

While $\Lambda_{\min}$ was determined from the data, it is instructive to
examine its theoretical implications for the minimal values of the parameters
$I$, $r$ and $\epsilon$. From eq.\ \ref{eq:J} and fig.\ \ref{fig:Ip}, the
smallest $I$ is \about\ 0.02, implying that the minimal values of the two other
parameters obey $r_{\min}\epsilon_{\min} \simeq 2\x\E{-3}$. The smallest $r$ is
obtained from the largest $\gamma$, the fraction of the mass carried away by
the BLR outflow (\S\ref{sec:r}). Since the outflow mass applies a back torque
on the underlying disk, the amount of material the wind can extract from the
disk is limited. In the case of magnetically driven disk winds, an estimate of
this upper limit based on the \cite{Blandford_Payne} self-similar solution
yields $\gamma \la 0.3$ \citep{Emmering92, Pelletier92}. This implies
$r_{\min}$ \about\,2 (eq.\ \ref{eq:r-gamma}) and $\epsilon_{\min}$ \about\,
\E{-3}. Mass extraction by the outflow might reach a higher maximum as a result
of other effects, e.g., radiation pressure. Increasing the maximal $\gamma$ to
50\%, so that fully a half of the mass entering the disk at the BLR outer
boundary is blown away and only the remaining half is eventually accreted by
the BH, yields $r_{\min}$ = 1 and $\epsilon_{\min}$ \about\,2\x\E{-3}. These
estimates for $\epsilon_{\min}$ are compatible with theoretical expectations
for low accretion rates: As \Mdot\ decreases below the values assumed in
classical accretion-disk solutions (\S\ref{sec:epsilon}), thermalization times
become longer, the solution switches to the advection dominated accretion flow
(ADAF) mode and the radiative efficiency decreases to \about\ \E{-3}
\citetext{for recent reviews, see \citealt{Narayan02}; \citealp{Yuan07}}.


\begin{table*}
\centering
\def\s{\phantom{0}}
\begin{tabular}{lccccrc}
  \hline  \hline
     \s\s(1)    &   (2)   &   (3)   &   (4)   &   (5)   & (6)\s& (7)  \\
      Source ID &   $L$   &   $M$   &  $\ell$ &  \Lmm   &\emm\s& ref  \\
                & (\erg)  &  (\Mo)  &         & (\erg)  & (\%) &      \\
    \hline \hline

 IRAS01428-0404 & 9.8E+41 & 2.0E+07 & 3.9E-04 & 5.6E+44 &  0.3 & S,SL \\
       NGC 3147 & 5.0E+42 & 4.1E+08 & 9.7E-05 & 4.2E+45 &  0.2 &    B \\
     J1231+1106 & 5.7E+42 & 1.0E+05 & 4.5E-01 & 1.6E+43 & 13.6 &    H \\
    3XMM 275370 & 9.5E+42 & 3.7E+08 & 2.0E-04 & 3.9E+45 &  0.3 &    P \\
     3XMM 47793 & 1.1E+43 & 8.6E+08 & 9.7E-05 & 6.8E+45 &  0.2 &    P \\
       NGC 3660 & 1.5E+43 & 7.1E+06 & 1.7E-02 & 2.8E+44 &  3.3 &    B \\
  1ES 1927+6054 & 1.6E+43 & 2.2E+07 & 5.8E-03 & 5.9E+44 &  2.0 & T,SL \\
     3XMM 93640 & 1.9E+43 & 6.5E+07 & 2.3E-03 & 1.2E+45 &  1.3 &    P \\
    3XMM 266125 & 3.2E+43 & 8.4E+07 & 3.0E-03 & 1.5E+45 &  1.7 &    P \\
    3XMM 303293 & 3.7E+43 & 1.2E+09 & 2.4E-04 & 8.7E+45 &  0.5 &    P \\
    3XMM 305003 & 4.3E+43 & 5.6E+08 & 6.1E-04 & 5.2E+45 &  0.8 &    P \\
    3XMM 339379 & 5.6E+43 & 3.2E+08 & 1.4E-03 & 3.5E+45 &  1.4 &    P \\
      3XMM 1780 & 6.6E+43 & 1.2E+09 & 4.4E-04 & 8.5E+45 &  0.8 &    P \\
        GSN 069 & 8.0E+43 & 1.2E+06 & 5.3E-01 & 8.6E+43 & 28.5 &    M \\
      Q2131-427 & 2.5E+44 & 7.7E+08 & 2.6E-03 & 6.4E+45 &  2.6 &    B \\

  \hline \hline

\end{tabular}

\caption{Confirmed true type 2 AGN: (1) Source identification; (2) bolometric
luminosity; (3) black-hole mass; (4) Eddington ratio $\ell = L/\Ledd$; (5) the
largest luminosity and (6) smallest radiative efficiency that would still be
consistent with a true type 2 classification for the source (from eqs.\
\ref{eq:Lmm} and \ref{eq:em}, respectively); (7) references: B -
\citet{Bianchi12b}; H - \citet{Ho12}; M - \citet{Miniutti13}; P -
\citet{Pons14}; T - \citet{Tran11}; S - \citet{Shi10}; SL - \citet{Stern12b}}
\label{table}

\end{table*}

\subsection{High-luminosity true type 2 AGN}

Figure \ref{fig:L-M} summarizes our results, showing in the $L-M$ plane the
luminosity boundaries for observable broad emission lines for the two extremes
of $\Lambda$ (eqs.\ \ref{eq:Lam_min}, \ref{eq:Lam_max}). Hatching indicates the
domain of super-Eddington luminosities. Because of the different
$M$-dependences of \Lmin\ and \Ledd, their contours intersect in the $L-M$
plane at a mass whose value can be determined from eq.\ \ref{eq:Lmin-Edd}. The
lower red boundary, marked ``min $\Lambda$", corresponds to the minimal values
of $I$, $r$ and $\epsilon$. As the parameters increase, the boundary moves
upwards and when all three reach their maximal values, $\Lambda$ is the largest
and the boundary becomes the upper red curve, marked ``max~$\Lambda$". If broad
line emission is properly described by the disk-wind scenario then \emph{all
AGN below the low boundary are true type 2 and all AGN above the upper boundary
are broad line emitters}. We reemphasize that this prediction is an immediate
consequence of mass conservation and applies to all disk outflows whether
smooth or clumpy and whatever the dynamics model at their foundation.

The lower red boundary corresponds to the luminosity \LEH\ (eq.\ \ref{eq:LEH}).
As noted above, this is currently the lowest luminosity observed for any
broad-line emitting AGN \citep{EH09}. But the disappearance of all broad line
emission below \LEH\ does not imply that it must exist in every object above
this limit: Sources above the lower red boundary \emph{can} be true type 2,
below the upper red boundary they \emph{can} be broad-line emitters, hence in
between these boundaries they can be either. An AGN in this intermediate region
will display broad line emission only if its own parameters are such that
$L/M_7^{2/3} > \Lambda(I, r, \epsilon)$, otherwise it will be a true type 2. In
particular, just below the upper red boundary an AGN could still be a true type
2 if it had a standard thin disk around a maximally spinning black hole
($\epsilon = 0.3$) and if its $r$ and $I$ were at the upper end of their
ranges. We will refer to objects with $L/M_7^{2/3} < \Lambda(I, r, \epsilon)$
in the $L-M$ domain between the two red boundaries as \emph{high-luminosity
true type 2 AGN}. Since the BLR can disappear anywhere below the upper red
boundary, {\em type 2 AGN can exist at any Eddington ratio and at luminosities
as high as a few \E{46} \erg}. Sources that have both high masses and high
Eddington ratios will not become true type 2---the Eddington ratio of a
\E{10}\,\Mo\ AGN on the upper red boundary is only 2.8\%. Searches for true
type 2 AGN will have higher success rates at lower Eddington ratios because
every AGN can be a true type 2 at Eddington ratio below \about\,1\%. Similarly,
at masses lower than 2.2\x\E5\,\Mo\ every sub-Eddington AGN can be a true type
2.

\section{Comparison with Observations}
\label{sec:true2}

To determine whether broad lines are missing from the observed spectrum of an
AGN we need to establish the line intensity expected in the source. From the
large body of observations, empirical relations have been derived for the
dependences of line equivalent widths (EWs) on source luminosity, host galaxy
luminosity, the aperture used for the spectroscopic observations, and the
signal-to-noise of the observations. Perhaps the most suitable line in the
optical part of the spectrum is \Ha, the strongest broad emission line observed
from the ground in low redshift AGN. Detailed studies of broad emission lines
in low redshift SDSS AGN of low to intermediate luminosity show a large range
of 300--600\A\ for EW(b\Ha), the broad \Ha\ equivalent width relative to the
intrinsic AGN continuum \citep[and references therein]{Greene05, Stern12a}. The
distribution is very broad and sources with EW(b\Ha) smaller than 200\A\ are
quite common. Very high luminosity type 1 AGN tend to show somewhat smaller
EW(b\Ha) \citep[e.g.][]{Capellupo15}, but the difference may not be
statistically significant. Most, perhaps all of these sources are expected to
be powered by standard, high efficiency accretion disks \citep{Netzer14}. The
observational uncertainties on EW(b\Ha) can be large, especially among low
luminosity AGN where the host galaxy contribution inside the chosen aperture
(e.g. the 3\arcsec\ SDSS fiber) can exceed the AGN contribution, making the
subtraction of stellar light quite uncertain.  The uncertainties for type 1
LINERs, which may be powered by ADAF, are so large that it is not at all clear
whether the typical EW(b\Ha) distribution of such sources is similar to that
observed in the other type 1 AGN. The mean observed value adopted here is
EW(b\Ha) = 500\A.

According to \cite{Stern12a}, there is a strong linear correlation between the
broad \Ha\ luminosity, \LbHa, and the AGN bolometric luminosity $L$, as derived
from X-ray observations, and the above mean EW corresponds to $L \simeq 130
\LbHa$. Practical lower limits for \Ha\ detection in type 1 AGN can be set at
10\% of the expected values, namely, EW(b\Ha) $\simeq$ 50\A\ or \LbHa\ $\simeq
L/1300$. Below these limits we classify the source as a true type 2 AGN.
Sources with intermediate EW(b\Ha), of order 50--100\A, are considered to be
intermediate-type AGN. We also note that EW(\Hb) $\simeq$ 0.2\,EW(\Ha)
typically, thus our critical EW(b\Ha) corresponds to EW(b\Hb) $\simeq$ 10\A;
this would be a very weak broad line, hard to detect among the numerous broad
and narrow lines in the wavelength range 4800--4900\A\ centered on \Hb.

It is not our intention to carry out a systematic search of the literature for
true type 2 AGN that obey the strict upper limit set here on the broad \Ha\
line. Instead we focus on a small group where a very detailed, high quality
study has already been carried out to establish the nature of each source as a
true type 2 AGN. Table 1 summarizes the properties of such published cases,
listing them in order of increasing $L$. The list is by no means complete but
rather based on carefully checked observations known to us. The listed
luminosities and BH masses are from the original papers. In the absence of
broad lines, the mass estimates are generally based on correlations with bulge
properties, which carry a typical error of 0.5 dex \citep[e.g.][]{Lasker14}.
The uncertainty in $L$ depends on the X-ray observations and the conversion
between $L$(2--10 keV) and $L$. The assumption for all sources listed here is
that their X-ray luminosity scales with $L$ the same as in ``typical" type 1
AGN, an assumption that is not directly testable for true type 2s. In this
case, the uncertainty on $L$ is about 0.2--0.3 dex. Of the 15 listed sources,
the two with the smallest BH masses stand out with their very high Eddington
ratios.

The quantities \Lmm\ and \emm, tabulated in columns 5 and 6, respectively, were
derived for each object from the disk-wind scenario described above. According
this scenario, an AGN transitions into the true type 2 class when its
luminosity drops below $\Lambda M_7^{2/3}$ (eq.\ \ref{eq:Lmin}). Denote by
\Lmm\ the transition luminosity corresponding to the maximal value of $\Lambda$
(eq.\ \ref{eq:Lam_max})
\eq{\label{eq:Lmm}
   \Lmm = 3.5\x\E{44}\,M_7^{2/3}\ \erg.
}
Listed in column 5, \Lmm\ would be the location on the upper red boundary in
Figure \ref{fig:L-M} of an AGN with the same BH mass. Given the uncertainty on
the BH mass, the uncertainty on \Lmm\ is about 0.3 dex. For a given AGN,
absence of broad-line emission is compatible with the mass conservation bound
if its luminosity $L$ is lower than its \Lmm. Since every listed source has $L
< \Lmm$, they all comply with the disk-wind constraint on true type 2 AGN. In
most cases, the bound is obeyed by a sufficiently comfortable margin that
\Lmin\ need not be as large as \Lmm, implying that $\Lambda$ need not be at its
maximum, i.e., the parameters $I$, $r$ and $\epsilon$ can be smaller than their
maximal values, in particular $\epsilon$ could be less than 0.3 (eq.\
\ref{eq:epsilon}).

The smallest radiative efficiency, \emm, that a source could have and still
comply with the condition for BLR disappearance is the one that produces $\Lmin
= L$ with $r$ and $I$ at their respective maxima. From eq.\ \ref{eq:Lmin},
$\Lmin \propto \epsilon^{4/3}$ when $r$ and $I$ are held fixed, therefore
\eq{\label{eq:em}
   \emm = 0.3\,\left(\frac{L}{\Lmm}\right)^{3/4},
}
the quantity entered in column 6 of Table 1. This is an estimate of the minimal
$\epsilon$; the actual radiative efficiency of the source can be anywhere
between \emm\ and 0.3. The uncertainty on \emm\ depends on the uncertainties on
$L$ and $M$, and is of order 0.4 dex. The derived range of $\epsilon$ is
reasonably tight for the two objects with the highest Eddington ratios,
consistent with thin accretion disks with a spin parameter larger than 0.7,
i.e. close to maximum spinning. For all other sources the range is too wide for
a meaningful constraint. Still, it is interesting that they all have \emm\
below the thin disk minimum  of 0.038, consistent with the ADAF disks expected
at their low Eddington ratios, which are below \about 1\% .

Figure \ref{fig:L-M} shows the positions of the tabulated sources in the $L-M$
plane. All of them fall below the upper red boundary, in agreement with a true
type 2 classification for values of $\Lambda$ intermediate between its two
extremes (eqs.\ \ref{eq:Lam_min}, \ref{eq:Lam_max}). The absence of broad line
emission in all of these unobscured sources agrees with the luminosity
constraint arising from mass conservation in the disk-wind scenario.

\section{Summary and Discussion}

Outflows are widely recognized as an important component of the AGN
environment, and magnetically driven winds are plausible candidates for the
origin of high-velocity outflows from the inner radii of AGN accretion disks
(\citealt{Peterson06, SloneNetzer12}, and references therein). Broad line
disappearance at some low AGN luminosity is inherent to the BLR outflow
scenario since broad line emission requires a minimal column density, implying
a minimal outflow rate and thus a minimal accretion rate. This result is
independent of the detailed wind dynamics, which is still poorly understood.
Although a general formulation of the disk outflow from first principles does
not yet exist, the luminosity bound formulated here is applicable whatever the
details of such future theory might be. Dynamics considerations may only add
additional constraints. For example, \cite{Nicastro00} considered the interplay
between radiation- and gas-pressure in disk outflows and found a different
bound on the luminosity of broad-line emitting AGN. This model-specific bound,
and others like it, can only supplement the mass-conservation bound derived
here, which is always applicable.

The fundamental constraint on broad line emission conveniently splits the
problem into its separate elements. The minimal column density \Nmin\ required
for broad line emission is determined purely by atomic processes, independent
of any considerations of the BLR detailed structure. The black-hole and outflow
parameters combine to form another characteristic column, \Nw. Mass
conservation imposes the limit in eq.\ \ref{eq:Nw} that involves these two
fundamentally different column densities and the additional parameter $I$, the
only quantity with dependence on the wind detailed structure (eq.\
\ref{eq:Mw}). Because \Nmin\ is different for different emission lines, various
broad lines could disappear at somewhat different luminosities and produce line
ratios that differ from standard broad line spectra. For example, the main
emission region for the Balmer lines is about three times further away than the
formation region of \ion{C}{IV}1549. The latter also require smaller critical
column. Thus a reduction in accretion rate could produce an abnormally large
(relative to type 1 AGN) \ion{C}{IV}1549/H$\alpha$ line ratio during the
transition stage from type 1 to true type 2.

The resulting limit on luminosity of broad-line emitters (eq.\ \ref{eq:Lmin0})
is independent of the wind dynamics. The large spread in luminosities of the
transition to true type 2 (eqs.\ \ref{eq:Lam_min}, \ref{eq:Lam_max}) primarily
reflects the large range {predicted for the} radiative efficiency $\epsilon$,
covering the entire range of fast spinning BHs surrounded by thin accretion
disks to low efficiency ADAFs. When operating at the upper end of the range of
$\Lambda$, an AGN with $M = \E{10}\Mo$ will become a true type 2 when its
luminosity drops below 3.5\x\E{46}\,\erg, well inside QSO range. Objects with
high efficiency are those where the same $L$ is obtained with lower \Mw\ and
hence are more likely to lose their broad line emission, especially if the
outflow also has a shallow radial density profile. This scenario predicts that
true type 2 AGN should show an increased abundance of BH with high spin
parameter.

An AGN loses its broad line emission when $L/\Lmin$ decreases to $< 1$, which
can happen at any Eddington ratio (see eq.\ \ref{eq:Lmin-Edd} and fig.\
\ref{fig:L-M}). The present analysis shows that the AGN sub-pc structure could
be controlled by an entirely different luminosity scale, $\Lmin = \Lambda
M_7^{2/3}$ (eq.\ \ref{eq:Lmin}). The scale factor $\Lambda$ not only varies
among AGN but for a given source it can also vary as it evolves. Does every AGN
go through a true type 2 phase? We note that $L \propto \epsilon\,\Mdot$ while
$\Lmin \propto \epsilon^{4/3}$ when the other properties remain constant. The
AGN will lose its BLR as \Mdot\ is decreasing only if $L/\Lmin \propto
\Mdot/\epsilon^{1/3}$ drops below unity at some point; that is, the transition
to a true type 2 occurs only if $\Mdot/\epsilon^{1/3}$ decreases with the
accretion rate, therefore the issue revolves around the behavior of this ratio.
The fact that true type 2 AGN do exist indicates that this ratio does indeed
decrease with \Mdot, at least in some sources. For standard accretion disks,
$\epsilon$ cannot change on a short (several thousand years) time scale, since
this is basically the BH spin, and the transition to true type 2 follows
directly from the drop of \Mdot.

\subsection{Toroidal Obscuration Region}

Although our focus here is BLR properties, the disk-wind scenario has similar
implications also for the TOR. However, deriving a quantitative estimate for
the equivalent of \Lmin\ (eq.\ \ref{eq:Lmin}) for TOR disappearance is more
difficult because it brings in the outflow properties of the TOR, whose
relation to those of the BLR and the AGN bolometric luminosity $L$ is unknown.
A direct comparison of the radial columns through the TOR and BLR can be
obtained with the aid of eq.\ \ref{eq:Mw}, which shows that \hbox{$\NR \propto
\Mw/I$} for the column through each of these two outflow regions, with the same
proportionality coefficient for both. Therefore \hbox{\NRT =
\NRB\x(\MwT/\MwB)\x(\IB/\IT),} with superscripts denoting the respective
quantities in each of the two regions. From the definition of $I$ in eq.\
\ref{eq:J} it is straightforward to show that $\IB/\IT < 1$, whatever the
functional form of the radial density profile $\eta$.\footnote{Because of the
normalization of $\eta$ (eq.\ \ref{eq:eta}), the integration in eq.\ \ref{eq:J}
obeys $\int\eta y^{1/2}dy = \ybar^{1/2}$ where \ybar\ is the value of $y$ at
some point inside the integration range. And because $y \ge 1$ for the TOR
while $y \le 1$ for the BLR, $J$ is $> 1$ for the former and $< 1$ for the
latter. This also shows that $J \to 1$ for either region when $Y \to 1$ (cf
fig.\ 1).} But the ratio \MwT/\MwB\ is entirely unknown, and cannot be
estimated without a full theoretical model for the disk outflow around AGN.
Such a model would have to include all relevant forces, in particular the
radiation pressure which can be widely different in the two regions.

Since a calculation of \Lmin\ is not yet feasible for the TOR, we cannot
determine in any given source whether the TOR should disappear before the BLR
or the other way round. In principle, there could exist both true type 2 AGN
and broad line emitters with and without dust obscuration. However, the basic
fact that the TOR must disappear below some accretion rate is inherent to the
disk wind scenario: Dust obscuration requires a minimal column density
therefore the TOR, too, must disappear at sufficiently low accretion rates and
with it the AGN mid-infrared (MIR) emission. That is, dust obscuration of both
the central continuum and the broad lines should disappear at some low
luminosity and the ratio $L_{\rm MIR}/L$ should decrease when the luminosity
drops below a certain, yet undetermined value.

{The above predictions are partly supported by the observations.
\cite{Chiaberge99} find that torus obscuration disappears in low-luminosity
($\la$~\E{42} \erg) FR~I radio galaxies, and \cite{Maoz05} find similar results
for LINERs. MIR observations are required to look for torus dust emission in
such systems. Some evidence supporting the decline of MIR emission in LINERs
comes from recent works by \cite{Rosario13} and \cite{Gonzalez15}.
Disappearance of the torus IR emission has been reported also in a number of
individual low-luminosity sources \citep{Whys04, Perlman07, MullerSanchez13}.}

The predicted torus disappearance at low $L$ does not imply that the disk wind
is abruptly extinguished, only that its outflow rate is lower than in
high-luminosity AGN. When the mass outflow rate drops below these ``standard
torus'' values, the outflow still provides toroidal obscuration as long as its
column exceeds \about\ \E{21} \cs. Indeed, \cite{Maoz05} found that some LINERs
do have obscuration, but much smaller than ``standard''. Line transmission
through a low-obscuration torus might also explain the low polarizations of
broad \Ha\ lines observed by \cite{Barth99} in some low luminosity systems.

{
\subsection{Potential Alternative Explanations}

Could there be other explanations for broad line disappearance? Since broad
lines arise from reprocessing of the central continuum, detectable broad
emission lines require sufficiently strong ionizing continuum and efficient
conversion of this continuum to line radiation. The latter involves the BLR
column density, i.e., the matter distribution in the radial direction, and its
covering factor, i.e., the angular distribution. Having concentrated on the
column density, we now examine the potential role of the two other factors.

\subsubsection{Disk SED}

We can estimate the influence of different SEDs on observed line EWs using
previous studies of large AGN samples. In particular, in sources powered by
standard accretion disks, we can examine the possible connection between line
EW and the intensity of the ``big blue bump" (BBB), the part of the SED around
1000--3000\A\ which is directly observed in many sources and must be related to
the shape of the ionizing continuum at wavelengths below 912\A.

Thin accretion disk models show that, given BH mass and spin, a smaller
accretion rate is more noticeable in the short wavelength part of the spectrum
and hence affects the Lyman continuum radiation more than the continuum at
4861\A\ or 6563\A, the respective wavelengths of the \Hb\ and \Ha\ lines. Such
variations also affect the intensity of the BBB relative to the longer
wavelength continuum. The shape of the SED is also associated with BH mass.
Accretion disks around more massive BHs but similar $L/\Ledd$ and BH spin are
predicted to have weaker BBB and smaller ratios of Lyman continuum to optical
luminosity, which would decrease the Balmer line EWs. All predictions regarding
disk SEDs are verified by a recent detailed comparison of accretion disk models
with the data of 39 AGN \citep{Capellupo15, Capellupo16}. Here, and in several
earlier, somewhat less detailed studies (see \citealt{Capellupo15} for
references), the steeper SED and weaker BBB as functions of accretion rate, BH
mass and spin are all demonstrated quite clearly. Given this, one may argue
that true type 2 AGN are weak line emitters because of their softer SED, the
result of the lower accretion rate. A definitive study of this issue requires
comparisons of line EWs in larger samples that cover sources with different
luminosity, accretion rate (or $L/\Ledd$) and BH mass, and represent the entire
AGN population (which the Capellupo et al sample did not).

The observational situation regarding broad line EWs has been studied in great
detail under the title  ``The Baldwin effect'' \citep{Baldwin77}, the well
known correlation of the EWs of several broad emission lines with continuum
luminosity. While clearly observed in several strong UV lines, such as
\ion{C}{IV}1549 \citep[][and references therein]{Risaliti11}, the hydrogen
Balmer lines produce conflicting results. For example, \cite{Greene05} provide
observational correlations between L(\Ha), L(\Hb) and $L_{5100}$ (the 5100\A\
luminosity) for sources with $L_{5100}$ = \E{42.5}--\E{45} \erg. For the \Hb\
line this study finds a correlation that can be translated to EW(\Hb) $\propto
L_{5100}^{1.13}$, the opposite of what is observed for \ion{C}{IV}1549, i.e. an
\emph{inverse} Baldwin effect. However, the systematic study by
\cite{Stern12a}, including sources with $L_{5100}$ = \E{41.8}--\E{45.2} \erg,
shows a ratio of \LbHa/$L_{5100}$ that does not vary with the source
luminosity. This can be interpreted as EW(b\Ha) which is independent of
luminosity, i.e., no Baldwin effect. As yet another example we use the sample
of 135 sources discussed by \cite{Netzer04}. In this case $L_{5100}$ ranges
from \E{43.4}--\E{47.5} \erg\ and EW(\Hb) varies from about 100--120\A\ at the
lowest luminosities to 60--70\A\ for the most luminous AGN in the universe,
i.e., a very weak Baldwin effect. This sample contains the most massive BHs,
yet the range in EW(\Hb), and $L/\Ledd$, is not very different from that in
sources where the BH mass, and AGN luminosity, are 3 orders of magnitude lower.
Finally, studies of very large samples, such as SDSS, also show EW(\Hb) \about\
60-100\A\ for $L_{5100}$ = \E{43.5}--\E{45.5} \erg, and no correlation between
the two (see fig 7.16 in \citealt{Netzer13}).

It is not our intention to discuss the origin of the Baldwin relationship---an
area of much confusion, dominated by various selection effects and biases with
no clear physical explanation. We simply point out that the observed Balmer
line EWs do not change much across a very large range of physical conditions,
much larger than that spanned by the 15 sources in Table 1. This is an
indication that the changing ionizing continuum of standard thin accretion
disks cannot by itself be  the origin of the changing line EWs in type 1 AGN.
The conclusion is that, much like type 1 AGN, true type 2 AGN can show weak or
strong BBB, depending on their other properties. Needless to say, the
predictions of the standard accretion disk scenario regarding the ionizing flux
cannot be extrapolated in a simple way to the lower efficiency ADAF systems,
where the theory is far less understood.

To explain objects that meet the criterion we set for true type 2 AGN
(\S\ref{sec:true2}) purely by SED effects, the number of ionizing photons would
have to be reduced by a full factor of 10. However, since all of the sources
listed in Table 1 show strong, high-ionization narrow lines, their broad line
deficits could be attributed purely to a change in the SED only if that change
had occurred within the past \about 1000 years, the typical recombination time
for the narrow lines region (NLR). A sharp drop in luminosity a few decades ago
is a possibility since the NLR cannot disappear on such a short time scale.
However, $L/\Ledd$ for several of the sources in question places them in the
domain of standard accretion disks, where a factor 10 reduction in the
accretion rate would not produce a similar decrease in EW(b\Ha). For example,
our calculations of disk SEDs for 1ES1927+6054 (see Table 1) using the
\cite{SloneNetzer12} code show that an increase in accretion rate by a factor
of 10 relative to the one listed in the table, combined with an assumed spin
parameter of 0.7 and no change in covering factor, results in an increase by a
factor of 2.5 in EW(b\Ha). This is far below the difference required to explain
the big change in EW(b\Ha) between a ``typical" type 1 source and a true type
2. A several year monitoring of the sources in Table 1 could provide a more
decisive answer to this question. Although current data do not provide
definitive evidence, it seems unlikely that in each of the Table 1 sources the
number of ionizing photons is less than 10\% of that in type 1 AGN with similar
optical continuum luminosity.

\subsubsection{Covering Factors}

Whatever the AGN properties, the production of detectable broad lines requires
part of its sky to be covered by material that captures a sufficient fraction
of the ionizing continuum. The \cite{Stern12a, Stern12b} study of a large
SDSS-selected sample of type 1 AGN shows that at low luminosity, most of them
actually appear as intermediate types (type 1.x), with a reduced ratio of
broad-to-narrow line strength. For the same data, \cite{EHT14} show that the
ratio of broad line to bolometric luminosity decreases along the spectral
sequence type 1.0 $\to$ 1.x $\to$ true type 2, indicating a gradual decline in
broad-line covering factor as the accretion rate is decreasing. Thus the
broad-line disappearance could be attributed to a diminishing covering factor
without the need to invoke a changing column density.

Elitzur et al.\ proposed that this spectral evolution arises naturally if the
wind is seeded with clouds that dominate its broad line emission. Remarkably,
the same quantity \Nw\ (eq.\ \ref{eq:Nw}) sets the scale not only for the mass
outflow rate but also for the dynamics of cloud motions: the clouds are
accelerated against the gravitational pull of the central black hole by the ram
pressure of the wind in which they are embedded, and the ratio of these
opposing forces on a cloud with column density \NH\ is controlled by $\NH/\Nw$.
While the details of this specific model are beyond the scope of the present
paper, the direct connection with the wind column density is most relevant to
our basic proposal. This model involves the additional assumption of
cloud-dominated broad-line emission and thus is not nearly as universal as the
current result.

\subsubsection{Conclusions}

Weakening and disappearing broad emission lines in type 1 AGN can result from a
changing disk SED, a drop in BLR covering factor or a decrease in ionized
column. All three can be related to a declining accretion rate through the disk
and it is difficult to disentangle their effects. Moreover, reflecting the
matter angular and radial distributions, respectively, the covering factor and
column density may be inherently related to each other. But whatever the other
effects, the relation $L > \Lmin$ (eq. \ref{eq:Lmin0}) stands out in its
universality as an absolute lower limit for detectable broad line emission from
disk outflows. As noted before, broad line disappearance below this limit
arises directly from mass conservation without any additional assumptions about
the wind clumpiness, structure or dynamics, and thus is a fundamental property
of the disk-wind scenario. Within this scenario, broad-line disappearance
triggered by a decrease of either ionizing continuum or covering factor will
only produce true type 2 AGN with luminosity \emph{higher} than \Lmin.

}

\subsection*{Acknowledgements}

We thank Estelle Pons for generous help with her data. Special thanks to Luis
Ho and the anonymous referee for their most useful comments on the manuscript.
Support by NASA (ME) and Israel Science Foundation grant 284/13 (HN) is
gratefully acknowledged.

\def\v{}


\label{lastpage}
\end{document}